\documentclass[onecolumn, 
prb,preprintnumbers,amsmath,amssymb,floatfix]{revtex4}

\usepackage[linktocpage,bookmarksopen,bookmarksnumbered]{hyperref}
\usepackage{graphicx}
\usepackage{dcolumn}

\usepackage{amsmath,graphics,epsfig,color,verbatim,ulem}

\begin{document}

\setlength{\pdfpageheight}{\paperheight}
\setlength{\pdfpagewidth}{\paperwidth}

\title{Spin dynamics and an orbital-antiphase pairing symmetry in iron-based superconductors}
\author{Z. P. Yin}
\email{yinzping@physics.rutgers.edu}
\author{K. Haule}
\author{G. Kotliar}
\affiliation{Department of Physics and Astronomy, Rutgers University, Piscataway, NJ 08854, United States.}
\date{November 1, 2013}
\maketitle

\textbf{
The symmetry of the wave function describing the Cooper pairs
is one of the most fundamental quantities in a superconductor
but its measurement in the iron-based superconductors has proved to be very difficult.
The complex multi-band nature of these materials makes the interplay
of superconductivity with spin and orbital dynamics very intriguing,
leading to very material dependent magnetic excitations, and pairing
symmetries.\cite{Stewart,Hirschfeld, PDai}
Here we use first-principles many-body method, including \textit{ab initio}
determined two-particle vertex function, to study the spin dynamics
and superconducting pairing symmetry in a large number of iron-based
superconductors.
In iron compounds with high transition temperature, we find both the
dispersive high-energy spin excitations, and very strong low energy
commensurate or nearly commensurate spin response, suggesting that
these low energy spin excitations play the dominate role in cooper
pairing.
We find three closely competing types of pairing symmetries, which
take a very simple form in the space of active iron $3d$ orbitals, and
differ only in the relative quantum mechanical phase of the $xz$, $yz$ and
$xy$ orbital contributions.
The extensively discussed s$^{+-}$ symmetry appears when contributions
from all orbitals have equal sign, while the opposite sign in $xz$ and
$yz$ orbitals leads to the $d$ wave symmetry. A novel orbital antiphase
$s^{+-}$ symmetry emerges when $xy$ orbital has opposite sign to $xz$ and $yz$
orbitals.
We propose that this orbital-antiphase pairing symmetry explains the
puzzling variation of the experimentally observed superconducting gaps
on all the Fermi surfaces of LiFeAs~\cite{LiFeAs, LiFeAs2}.  This
novel symmetry of the order parameter may be
realized in other iron superconductors.
}

The spin and the multi-orbital dynamics of iron based superconductors
is believed the play the essential role in the mechanism of
superconductivity~\cite{Mazin_PRL}, but a realistic modeling of
magnetic excitations, and a clear physical picture for their variation
across different families of iron superconductors, is currently
lacking. The Cooper pairs are locked into singlets but the orbital
structure of the superconducting order parameter can be material
dependent, and its connection to orbital and spin excitations is an
open problem.
To address these issues, we use \textit{ab initio} theoretical method
for correlated electron materials, based on combination of dynamical
mean field theory (DMFT) and density functional theory
(DFT)~\cite{review}. This computational method improves on DFT
description of electronic structure in iron superconductors, and
predicts correct magnitude of ordered magnetic moments~\cite{Yin-NM},
improves electronic spectral function and Fermi
surfaces~\cite{Yin-NM,Yin-PRB}, and charge response such as optical
conductivity~\cite{Yin-NP}. To successfully predict dynamical magnetic
response across different families of iron superconductors, and
superconducting pairing, it is crucial to compute from
\textit{ab initio} the two particle scattering amplitude also called
the two particle vertex function.\cite{HPark-PRL} Its complex calculation previously
precluded \textit{ab initio} description of trends across different
families of iron superconductors, as spin response is very sensitive
to the value and the energy dependence of this screened
interaction. In many methods used earlier, such as random phase
approximation, it is treated as a static adjustable
parameter~\cite{Kuroki, Kreisel}, or it was computed by summing very
limited subset of Feynman diagrams~\cite{Onari}. Within the functional
renormalization group~\cite{FWang} this screened interaction is
computed rather well, but the electronic structure is greatly
simplified, treating the Hubbard model for iron $d$-bands only, and
usually neglecting renormalization of electronic structure due to
electronic correlations. Similar simplification is made in strong-coupling
approaches.\cite{Qimiao}

\begin{figure}[bht]
\includegraphics[width=0.9\linewidth]{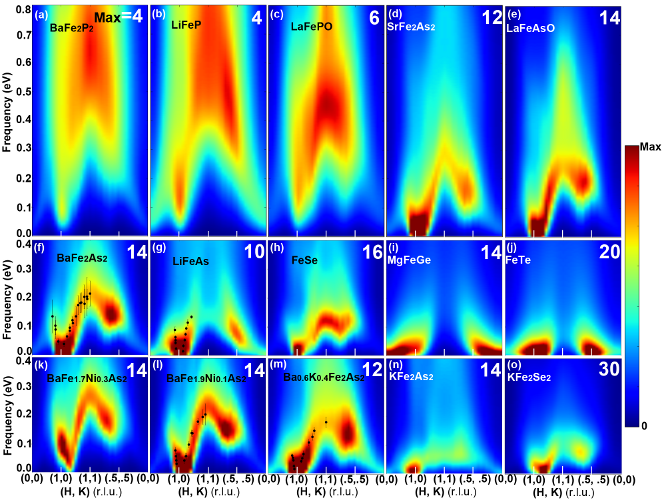}
\caption{ \textbf{Dynamic spin structure factor $S(q, \omega)$ in iron
    pnictides, chalcogenides and MgFeGe.}  The $S(q, \omega)$ is
  plotted along the high-symmetry path $(H,K,L=1)$ in the first Brillouin
  zone of the single iron unit cell. The intensity varies substantially across these
  compounds, hence the maximum value of the intensity was adjusted to
  emphasize the dispersion most clearly. The maximum value of the
  intensity in each compound is shown in the top right corner.
  The color coding corresponds to the theoretical calculations for
  (a) BaFe$_2$P$_2$ ($T_C^{max}<2K$);
  (b) LiFeP ($T_C=6K$); (c)LaFePO ($T_C=7K$); (d)
  SrFe$_2$As$_2$ ($T_C^{max}=37K$); (e) LaFeAsO ($T_C^{max}=43K$); (f)
  BaFe$_2$As$_2$ ($T_C^{max}=39K$); (g) LiFeAs ($T_C=18K$); (h)
  FeSe ($T_C^{max}=37K$); (i)MgFeGe ($T_C^{max}=0$); (j)FeTe ($T_C^{max}=0$); (k)
  BaFe$_{1.7}$Ni$_{0.3}$As$_2$ ($T_C<2K$); (l)
  BaFe$_{1.9}$Ni$_{0.1}$As$_2$ ($T_C=20K$); (m)
  Ba$_{0.6}$K$_{0.4}$Fe$_2$As$_2$ ($T_C=39K$); (n) KFe$_2$As$_2$ ($T_C=3.5K$); (o)
  KFe$_2$Se$_2$.  The experimental data are shown as black dots with error
  bars  in (f), (g), (l) and (m), digitized from Refs. \onlinecite{chi_BFA, chi_LiFeAs, chi_BFNA, chi_BKFA}.
}
\label{SQW-all}
\end{figure}
All iron-based superconductors contain similar layers of tetrahedra
with iron in the center and pnicogen/chalcogen at the corners,
but their spin excitation spectra varies greatly among compounds.
In Fig.~\ref{SQW-all} we plot the dynamic spin structure factor S($q$,
$\omega$)=$\chi''$($q$,$\omega$)/(1-exp(-$\hbar\omega$/k$_B$T)) for
several classes of iron compounds along the high symmetry momentum
path in the first Brillouin zone of the single iron unit
cell. We overlay the neutron scattering data~\cite{chi_BFA,
  chi_LiFeAs, chi_BFNA, chi_BKFA} for some compounds where experiment
is available, to show a good agreement between theory and
experiment. The spin-wave bandwidth, defined as the difference between
the minimum ($q=(1,0)$) and the maximum ($q=(1,1)$) of spin excitation
is related to spin-exchange $J$, which is inversely proportional to
the strength of the low energy Coulomb interaction ($J\propto t^2/U$),
hence increased correlation strength leads to smaller
bandwidth. Notice that the correlation strength in these compounds is
dominated by the Fe-pnicogen distance, as shown in Ref.~\onlinecite{Yin-NM}.
The phosphorus compounds (Figs.~\ref{SQW-all}a-c) show the largest spin
wave bandwidth of the order of $0.6\,$eV-$0.45\,$eV, which is a
consequence of their most itinerant nature among these
compounds~\cite{Yin-NM}.  The mass enhancement due to correlations is
increased in arsenides and even more in chalcogenides~\cite{Yin-NM},
hence the spin-wave bandwidth is progressively reduced to
$~0.3-0.2\,$eV in Figs.~\ref{SQW-all}d-f, and $~0.15\,$eV in
Figs.~\ref{SQW-all}g-h.
The intensity of spin excitation is
proportional to the size of the fluctuating moment in this energy range, 
which roughly anti-correlates with strength of correlations, hence phosphorus
compounds show the weakest ($Max=4$) and FeTe shows the strongest
($Max=20$) intensity.

\begin{figure}[bht]
\includegraphics[width=0.9\linewidth]{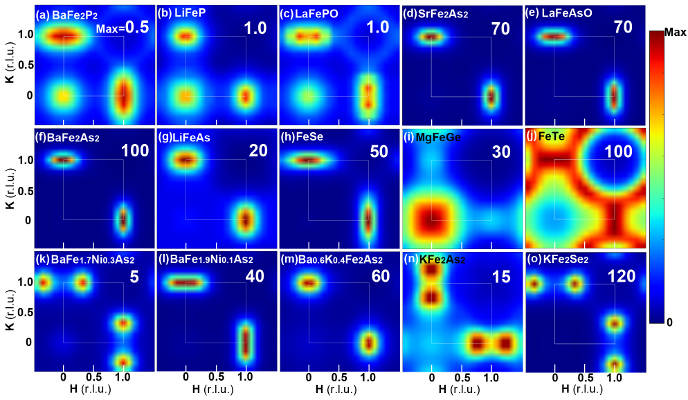}
\caption{ \textbf{Dynamic spin structure factor $S(q, \omega)$ in iron
    pnictides, chalcogenides and MgFeGe.}  The $S(q, \omega)$ is
  plotted in the 2D plane $(H,K)$ at constant $\omega$=5 meV for the same
  materials as in Fig.\ref{SQW-all}.  The maximum intensity scale for
  each compound is marked as a number in the top-right corner of
  each subplot. We take the cut at $L=1$ for all compounds except
  MgFeGe and phosphorus compounds, where $L=0$ plane is shown to
  emphasize their tendency towards ferromagnetism.
}
\label{SQW-2D}
\end{figure}
The low energy spin-excitations are much more sensitive to the details
of both the band-structure and the two-particle vertex function, hence
the trend across different compounds can not be guessed from either
the correlation strength or from the band structure. In
Fig.~\ref{SQW-2D} we show $S(q,\omega)$ for the same compounds as in
Fig.~\ref{SQW-all}, but we take a different cut. We keep the energy
fixed at $\omega=5\,$meV, and change momentum in the two dimensional
momentum plane $(H,K)$ (The momentum dependence in $z$ direction is
weak for most compounds).  As is clear from Figs.~\ref{SQW-all}a-c,
and Fig.~\ref{SQW-2D}a-c, the low energy spin-excitations are
extremely weak ($Max\approx 1$) in phosphorus compounds and the spin
excitations at the spin-density wave ordering vector $(1,0)$ is
comparable to its value at the ferromagnetic ordering vector
$(0,0)$. In strong contrast, the low energy spin excitations are very strong in
arsenides (Fig.~\ref{SQW-2D}d-g)
and are concentrated solely at the commensurate wave vector $(H,K)=(1,0)$.
This is the ordering wave vector of the spin-density wave magnetic
state, which is the ground state for these parent compounds, except
superconducting LiFeAs ($T_C=18\,$K).  When doped, all
compounds in Figs.~\ref{SQW-2D}d-f are high-temperature superconductors ($T_c\approx
37K-39K$). Similarly chalcogenide FeSe (Fig.~\ref{SQW-all}h), which
becomes superconducting at $T_c=37K$ under modest pressure $p=~3\,$GPa,
has similar low energy spin response as the arsenides superconductors.

On the other hand, MgFeGe is a compound with similar band structure as
LiFeAs, hence the random-phase approximation gives similar spin
response in the two compounds with low energy maximum intensity at
$(1,0)$.~\cite{HBRhee} Inclusion of the realistic two-particle vertex
function, as done in this study, has a profound impact on the spin
excitations. A broad maximum appears at $(0,0)$ (see
Fig.~\ref{SQW-2D}i), hence spin fluctuations are ferromagnetic in
agreement with calculation of Ref.~\onlinecite{MgFeGe-Mazin} showing
stable ferromagnetic ground state. Finally FeTe has also much broader
spin-excitations covering a large part of the Brillouin zone (see
Fig.~\ref{SQW-2D}j), and shows two competing excitations at $q$=(1,0)
and $q$=(0.5, 0.5), the latter corresponds to the ordering wave vector
of the low-temperature antiferromagnetic state of
Fe$_{1.07}$Te.\cite{moment-FeTe}

The common theme in high-temperature
superconductors (Figs.~d-h) is thus the existence of well defined high
energy dispersive spin excitations with bandwidth between
$0.1-0.35\,$eV, and most importantly very well developed commensurate
(or nearly commensurate) low energy spin excitations at wave vector
$q=(1,0)$, consistent with the theory of spin-fluctuation mediated
superconductivity~\cite{TMoriya, DPines}.
The pnictide parent compounds SrFe$_2$As$_2$, LaFeAsO, BaFe$_2$As$_2$
have strong low energy spin excitation centered exactly at $q=(1,0)$, while in LiFeAs and
FeSe the spin excitation is peaked slightly away from this commensurate
wave vector. Consequently, the former three compounds have
antiferromagnetic ground state, while the latter two are
superconducting.  
In the former, electron or
hole doping is needed to suppress the long range magnetic order, and
to stabilize the competing superconducting state. In
Figs.~\ref{SQW-all}f,k-n and Figs.~\ref{SQW-2D}f,k-n we illustrate the doping
dependence of the spin-excitation spectrum on the examples of electron
doped and hole doped BaFe$_2$As$_2$, i.e., BaFe$_{1-x}$Ni$_x$As$_2$ and
Ba$_{1-x}$K$_x$Fe$_2$As$_2$, respectively. The electron doping
slightly increases the bandwidth (comparing Fig.~\ref{SQW-all}(f) with
Fig.~\ref{SQW-all}(k)), whereas the hole doping dramatically reduces the
bandwidth from $\sim0.2\,$eV to $\sim0.05\,$eV in overdoped
KFe$_2$As$_2$\cite{chi_BKFA,Note-KFA} (Fig.~\ref{SQW-all}(n)). The low energy
spin excitations in the electron overdoped
BaFe$_{1.7}$Ni$_{0.3}$As$_2$ become very weak and strongly
incommensurate\cite{chi_BKFA} with peak centered at $q$=(1.0, 0.35) (see
Fig.~\ref{SQW-2D}k). Similarly, on the hole overdoped side in
KFe$_2$As$_2$, the low-energy spectrum is suppressed (maximum intensity
in Fig.~\ref{SQW-2D}n is 15 compared to 100 in the parent compound),
and main excitation peak moves to incommensurate $q$=(0.75, 0) in
agreement with experiment.\cite{KFA-SQW,chi_BKFA} The optimally doped
compounds (Figs.~\ref{SQW-all}l,m) have high energy spin excitations
very similar to the parent compound, while the low energy excitations
are slightly reduced and broadened in momentum space
(Fig.~\ref{SQW-2D}l,m), to suppress long range magnetic order of the
parent compound. This is very similar to the spectrum of LiFeAs and
FeSe, which both have superconducting ground state. From these plots,
we can deduce that near commensurate or commensurate spin excitations
at $q=(1,0)$, with some finite width in momentum space to reduce the
tendency towards the long-range order, are favorable for
superconductivity.

Now we comment on the complexity of the K$_x$Fe$_{2-y}$Se$_2$
compounds. Our results for KFe$_2$Se$_2$ in
Figs.~\ref{SQW-all}\&\ref{SQW-2D}(o) indicate strong low energy spin
excitation peaked around $q=(1,0.4)$.  Vacancies in the K site, which
reduce the effective electron doping, can move the peak towards $q=(1, 0)$ and
favor superconductivity. On the other hand, vacancies in the Fe sites
can move the peak to $q$=(0.6, 0.2) to induce novel magnetism in
K$_{0.8}$Fe$_{1.6}$Se$_2$~\cite{KFS-moment}.

Whereas the dynamic spin structure factor $S(q,\omega)$ dispersion and
the strength of the low energy spin excitations correlate with
experimental $T_c$ across many families of iron superconductors, the
superconducting pairing symmetry and the variation of the
superconducting gaps on the different Fermi surfaces cannot be
extracted from the spin dynamics alone. To make further progress on
these issues, we computed the complete two particle scattering
amplitude in the particle-particle channel within the dynamical mean
field theory framework, and we solved Eliashberg equations in the BCS
low energy approximation (see Supplementary material). In the
Eliashberg equations, the orbital degrees of freedom play the central
role, rather than the bands, because the Coulomb interaction, and the
two particle irreducible vertex function in the particle-hole channel,
is large between the iron-3$d$ electrons on the same iron site.

\begin{figure}[bht]
\includegraphics[width=0.9\linewidth]{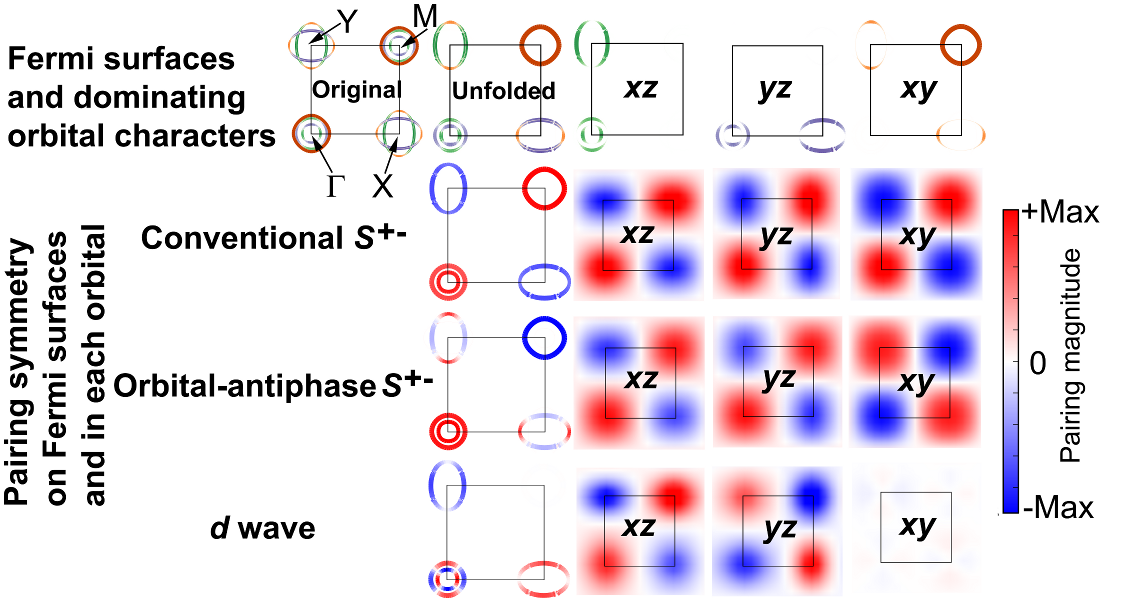}
\caption{
\textbf{Fermi surfaces, pairing symmetries and the basic building blocks.}
Top row: the original and unfolded two-dimensional Fermi surfaces
in the $\Gamma$ plane for the representative compound LaFeAsO in
paramagnetic state, shown in the first Brillouin zone of the single iron unit
cell.
On the top right, the Fermi surfaces are further decomposed into the dominating Fe-$t_{2g}$
($xz$, $yz$, and $xy$) characters.
The next three rows, from top to bottom, show respectively
the conventional $s^{+-}$, the orbital-antiphase $s^{+-}$ and the $d$-wave pairing symmetries
of the superconducting order parameter.
The left column shows the Fermi surfaces colored with strength of the
order parameter $\Delta_j(k)=<c^+_{k\uparrow,j}c^+_{-k\downarrow,j}>$ ($j$ is the band
index), while the right columns decompose the order parameter in orbital
space, i.e., $\Delta_{\alpha}(k)=<c^+_{k\uparrow,\alpha}c^+_{-k\downarrow,\alpha}>$
($\alpha$ runs over Fe-$t_{2g}$ orbitals: $xz$, $yz$, and $xy$).
}
\label{FS-pairing}
\end{figure}
In all arsenide and chalcogenide compounds with strong (nearly)
commensurate low energy spin excitation, we find that the Eliashberg
equations give three almost degenerate solutions with the largest
eigenvalues. The corresponding three eigenvectors, which are
proportional to the superconducting order parameter
$\Delta_{\alpha,\alpha'}(k)$, are almost diagonal in orbital space
($\alpha\alpha'$), and we denote the diagonal terms as
$\Delta_{\alpha}(k)=\Delta_{\alpha,\alpha}(k)$.  As shown in
Fig.~\ref{FS-pairing}, these three states are similar in nature, since
each orbital has sign-changing $s^{+-}$ structure in momentum space,
described by the formula
\begin{equation}
\Delta_{\alpha}(kx, ky)=\Delta_{nnn, \alpha} cos(kx)cos(ky)+\Delta_{nn, \alpha} (cok(kx)+cos(ky))/2
\end{equation}
but the phases on different orbitals are different.  Here
$\Delta_{nn}$ correspond to the nearest-neighbor and $\Delta_{nnn}$ to
the next-nearest neighbor pairing, and we find that $|\Delta_{nn,
  \alpha}|<<|\Delta_{nnn, \alpha}|$, hence we have predominantly
next-nearest neighbor pairing.  In Fig.~\ref{FS-pairing} we plot Fermi
surfaces and gap function in the first Brillouin zone of the single iron unit
cell. We plot gap function for $t_{2g}$
orbitals only, because gaps on the $e_g$ orbitals are much smaller and
their weight at the Fermi level is also small.  Although the symmetry
of pairing expressed in orbital space is $s^{+-}$ in all three states,
the projection to the Fermi surface leads to different global
symmetries. When all three $t_{2g}$ orbitals have the same phase, we
recover the conventional $s^{+-}$ state~\cite{Mazin_PRL}. If the $xz$
orbital has the opposite phase to the $yz$ orbital, the global
symmetry is of $d$-wave type. In this case the $xy$ orbital shows
negligible pairing. Finally, we find a novel type of state in which
$xz$ and $yz$ orbitals are in-phase, but the $xy$ orbital has the
opposite phase. We call this state the \textbf{\textit{orbital-antiphase $s^{+-}$
    state}}.

\begin{figure}[bht]
\includegraphics[width=0.9\linewidth]{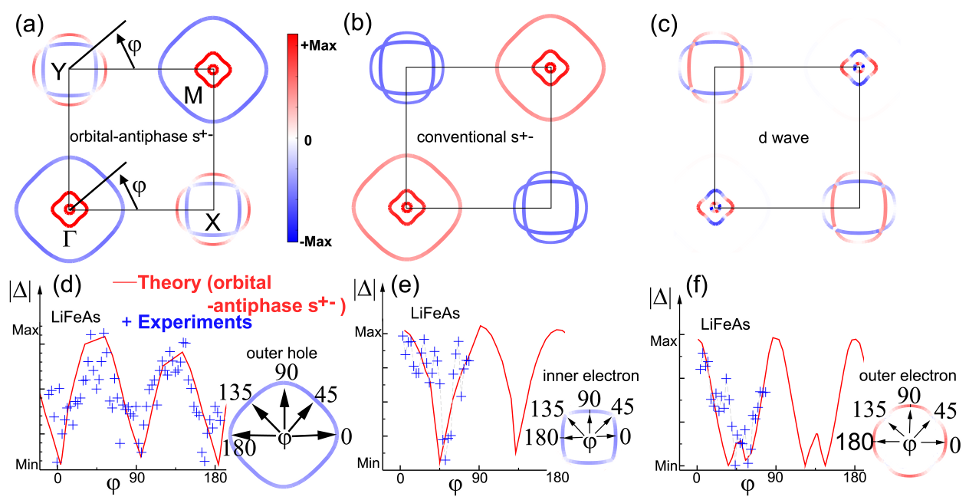}
\caption{
\textbf{Superconducting pairing symmetry and pairing amplitude anisotropy in LiFeAs.}
The superconducting order parameter $\Delta_j(k)=<c^+_{k\uparrow,j}c^+_{-k\downarrow,j}>$ on the Fermi surface,
with the (a) orbital-antiphase $s^{+-}$, (b) conventional $s^{+-}$ and
(c) a d-wave symmetry.
(d)-(f) show the variation of the pairing amplitude for the orbital-antiphase $s^{+-}$ state
on the outer hole Fermi surface and the two electron Fermi surfaces,
respectively. The red lines correspond to theoretical results, and the
$+$ symbols denote the experimental measurements from Ref.\onlinecite{LiFeAs}.
}
\label{LiFeAs-SC-gap}
\end{figure}
The identification of these basic building blocks greatly simplifies
our understanding of the superconducting pairing symmetry in the
iron-based superconductors, and explains why the superconducting
pairing symmetry is so sensitive to the details of the electronic
structure.
Since flipping the sign of the order parameter in a single
orbital has only a tiny energy cost, the first three eigenstates
discussed above are very close in energy and have very similar
eigenvalues, i.e., superconducting pairing strength.
In Fig.~\ref{LiFeAs-SC-gap} we show detailed results for LiFeAs, and
we plot all the three leading pairing symmetries on the two-dimensional 
Fermi surfaces in the $\Gamma$ plane of the tetragonal
crystallographic unit cell.
Our calculations show that the orbital-antiphase $s^{+-}$ has the
largest pairing strength in LiFeAs.
Experimentally, it was found\cite{LiFeAs, LiFeAs2} that the
superconducting gaps have very unusual variations on the Fermi
surfaces, where the superconducting gap is maximal (minimum) around
$\varphi$=45$^{\circ}$ (0$^{\circ}$) on the outer hole Fermi surface
(see Fig.~\ref{LiFeAs-SC-gap}(d)), whereas it is maximal (minimum)
around $\varphi$=0$^{\circ}$ (45$^{\circ}$) on the two electron Fermi
surfaces (Fig.\ref{LiFeAs-SC-gap}(e),(f)).
This observation was used as an evidence against the spin fluctuation
mechanism~\cite{LiFeAs} of superconductivity, as it is not consistent
with the conventional $s^{+-}$ state. Indeed, our calculation show
that conventional $s^{+-}$ order parameter can not account for the gap size
variation in momentum space, however, as seen from
(Fig.\ref{LiFeAs-SC-gap}(d)-(f)), the orbital-antiphase $s^{+-}$ can account
for the variation of this gap.

The novel orbital-antiphase $s^{+-}$ pairing symmetry is not limited
to LiFeAs but may be the pairing symmetry in many other iron-based
superconductors, such as Ba$_{1-x}$K$_x$Fe$_2$As$_2$ and
K$_x$Fe$_2$Se$_2$~[\onlinecite{Hong-Ding}]. Our study suggests that it
is important to describe the cooper-pairing in orbital-space, keeping
the complexity of orbital and spin fluctuations, which arise due to
electron correlations, rather than solving BCS equations for weakly
correlated systems in band space.

\textbf{Acknowledgments:} We thank H.~Park, P.~Dai and H.~Ding for stimulating
discussion. This work is supported by NSF DMR--1308141 (Z.P.Y. and G.K.) and
NSF Career DMR--0746395 (K.H.).

\clearpage

\setcounter{figure}{0}
\makeatletter
\renewcommand{\thefigure}{S\@arabic\c@figure}
\makeatother

\setcounter{equation}{0}
\makeatletter
\renewcommand{\theequation}{S\@arabic\c@equation}
\makeatother

\setcounter{table}{0}
\makeatletter
\renewcommand{\thetable}{S\@arabic\c@table}
\makeatother

\makeatletter
\renewcommand{\@biblabel}[1]{[S#1]}
\makeatother

{\centering
\textbf{Spin dynamics and an orbital-antiphase pairing symmetry in iron-based superconductors: Supplementary information}\\
}

\vskip 5mm

{\centering
Z. P. Yin,
\email{yinzping@physics.rutgers.edu}
K. Haule,
and G. Kotliar\\
}

{\centering
\textit{
Department of Physics and Astronomy, Rutgers University, Piscataway, New Jersey 08854, United States.\\
}
}

\vskip 5mm

\section{Method}

To carry out our first principles calculations taking into accounts strong correlation effects  
in the iron-based superconductors, 
we used a combination of density functional theory and dynamical mean
field theory (DFT+DMFT)~[S\onlinecite{DMFT-RMP2006}] as implemented in
Ref.~S\onlinecite{Haule-DMFT}, which is based on the full-potential
linear augmented plane wave method implemented in Wien2K~[S\onlinecite{wien2k}], 
together with the full two-particle vertex correction in the same footing~[S\onlinecite{HPark-PRLS}]. 
The electronic charge is computed self-consistently on DFT+DMFT density matrix. 
The quantum impurity problem is solved by the continuous time quantum Monte Carlo
method~[S\onlinecite{Haule-QMC,Werner}], using Slater form of the Coulomb repulsion in
its fully rotational invariant form. 
The local two-particle vertex is then sampled by the continuous time quantum Monte Carlo
method on the fully-converged DFT+DMFT solution.

We use the same experimentally determined lattice structures as in Ref.~S\onlinecite{Yin-NMS}, including the
internal positions of the atoms (see Table \ref{lattice-structure}), from 
References~[S\onlinecite{a-FeTe,a-FeTe-2,Guo,a-FeSe,a-LiFeAs,a-K122,a-Ba122,a-LaFeAsO,a-Sr122,a-LaFePO}].
We use the paramagnetic tetragonal lattice structures, and neglect the
weak structural distortions. For convenience, we include the Table S1 in Ref.~S\onlinecite{Yin-NMS}. 

\begin{table}[htb]
\caption{
Lattice constants and atomic positions of the compounds we studied.
$z1$ refers to the $z$ coordinate of P/As/Se/Te 
and $z2$ refers to coordinate of the other metallic atom in the 111 and 1111 structures.  
}
\label{lattice-structure}
\begin{tabular}{|c|c|c|c|c|c|c|}
\hline
Compounds       & $a$ ($\AA$)  & $c$ ($\AA$) & $z$(Fe) &  $z1$    &  $z2$     &  Ref.   \\
\hline
FeTe            & 3.8219       &  6.2851     & 0    & 0.2792    &           & S\onlinecite{a-FeTe, a-FeTe-2} \\
KFe$_2$Se$_2$   & 3.9136       & 14.0367     & 0.25 & 0.3539    &           & S\onlinecite{Guo}    \\
FeSe            & 3.7724       &  5.5217     & 0    & 0.2673    &           & S\onlinecite{a-FeSe}  \\
LiFeAs          & 3.7914       &  6.3639     & 0.5  & 0.7365    & 0.1541    & S\onlinecite{a-LiFeAs}  \\
KFe$_2$As$_2$   & 3.84         & 13.85       & 0.25 & 0.3531    &           & S\onlinecite{a-K122} \\
BaFe$_2$As$_2$  & 3.9625       & 13.0168     & 0.25 & 0.3545    &           & S\onlinecite{a-Ba122} \\
LaFeAsO         & 4.03533      &  8.7409     & 0.5  & 0.6512    & 0.14154   & S\onlinecite{a-LaFeAsO} \\
SrFe$_2$As$_2$  & 3.9243       & 12.3644     & 0.25 & 0.3600    &           & S\onlinecite{a-Sr122} \\
LaFePO          & 3.96358      & 8.51222     & 0.5  & 0.6339    & 0.1487    & S\onlinecite{a-LaFePO} \\
LiFeP           & 3.69239      & 6.03081     & 0.5  & 0.720055  & 0.151962  & S\onlinecite{a-LiFeP} \\
BaFe$_2$P$_2$   & 3.84         & 12.44       & 0.25 & 0.3456    &           & S\onlinecite{a-BaFe2P2} \\
\hline
\end{tabular}
\end{table}

We studied the paramagnetic phase of all compounds at the same temperature 
and used the same Coulomb interaction $U$ and Hund's coupling $J$ as in our previous work.[S\onlinecite{Yin-NMS}, S\onlinecite{Kutepov}, S\onlinecite{Yin-np}]

\subsection{Unfolding}

Since particle-hole irreducible vertices in our approach are nonzero
only for electrons in the $3d$ orbitals of the iron atoms, we can
rewrite the perturbation theory in terms of vertices defined in this
$3d$ orbital subspace, and the single particle Green's function
$G_{dd}(k,\omega)$, which starts and ends in the iron $3d$
manifold. This Green's function can be cast into the form, where we
consider only one iron per unit cell in larger Brillouin
zone. This operation is usually called \textit{unfolding}. Note that
here we keep all the information from the two-iron atom calculation
(for example, we have shadow bands in our Green's function), we merely
rewrite perturbation theory in terms of more convenient larger
Brillouin zone and one atom unit cell. We checked that for our choice
of unfolding, the off-diagonal terms of the single particle Green's
function in momentum $G_{dd}(k,k+Q)$ is negligible.

To transform the electron structures from the two-iron/unit cell to
the one-iron/unit cell, we used the following transformation:
\begin{equation}
G_{\alpha, \alpha'}(k)=\sum_{R, R'} e^{ik\cdot (R-R')} G_{R\alpha, R'\alpha'}(k)
\end{equation}
where $R$ and $R'$ are the positions of the two different Fe atoms in the original crystal structure, 
and $R'$ is related to $R$ by a group symmetry operation of the
crystal structure. $\alpha$,$\alpha'$ labels orbitals. Special care has to be taken to the phases of the
orbitals chosen on the two iron atoms. To make off-diagonal terms in
the green's function $G_{dd}(k,k+Q)$ negligible, we need to choose
different phases for the orbitals on the two atoms.

For example, in the case of BaFe$_2$As$_2$, we choose $R=(0.5,0,0.25)$
and $R'=(0.5,0,-0.25)$, and phases for all $3d$ orbitals on the second
atom are opposite to the first atom.
For LiFeAs and LaFeAsO, we choose $R=(0.75,0.25,0.5)$ and
$R'=(-0.75,-0.25,-0.5)$, consequently the phase are opposite for
Fe-$d_{z^2}$,$d_{x^2-y^2}$,$d_{xy}$ and equal for $d_{xz}$ and $d_{yz}$ orbitals.

\subsection{Magnetic and Charge Susceptibility}

\begin{figure}[bht]
\includegraphics[width=0.3\linewidth]{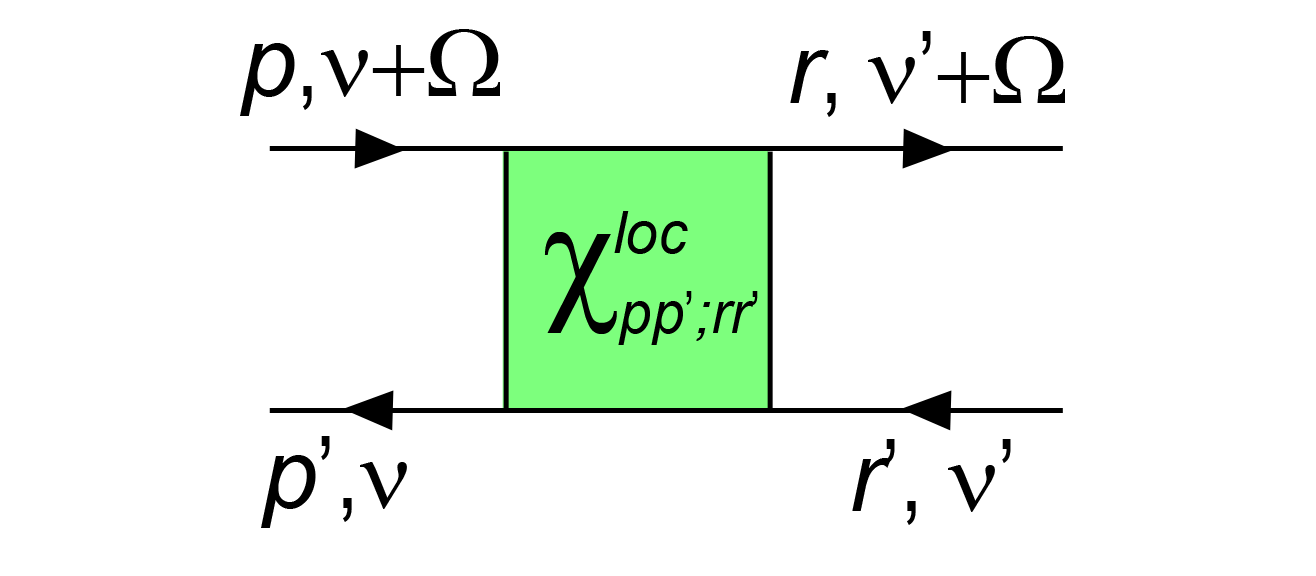}
\caption{
\textbf{Feynman diagram for the local two-particle susceptibility.}
}
\label{Chi}
\end{figure}
The general two-particle susceptibility (depicted in Fig.\ref{Chi}) is defined by
\begin{equation}
\chi_{pp';rr'}^{loc}=< \Psi_r^+(\tau_1) \Psi_{p'}^+(\tau_2) \Psi_{r'}(\tau_3) \Psi_{p}(\tau_4) >
\end{equation}
where $p$, $p'$, $r$, $r'$ are combined index of spin and orbital,
$\Psi^+$ is the creation operator and $\Psi$ the annihilation
operator.

The central quantity of this approach is the irreducible two particle
vertex $\Gamma$, which consists of all Feynman diagrams, and can not
be separated into two parts by cutting any two propagators.
There are several types of irreducible vertices: i) the irreducible
vertex in the particle-hole channel consists of all Feynman diagrams,
which can not be separated into two parts by cutting any two
propagators going in opposite direction, i.e., cutting a particle and a
hole ii) the particle-particle irreducible vertex correspondingly contains
diagrams which are not separated into parts by cutting any two
propagators going in the same direction, iii) the fully irreducible
two-particle vertex contains all diagrams which can not be broken into
separate diagrams by cutting either particle-particle or particle-hole
pair of propagators.

Within the Dynamical Mean Field Theory (DMFT), the particle-hole
irreducible vertex $\Gamma$ is local, and it is equal to the impurity
vertex, which can be obtained from the solution of the quantum
impurity model. This can be proven with the same power-counting
arguments as the self-energy is proven to be local in the limit of
large coordination number.[S\onlinecite{DMFT-RMP2006}, S\onlinecite{review1996}]
To compute $\Gamma^{imp}$, we sample the two-particle susceptibility
$\chi^{imp}$ by the quantum Monte Carlo impurity solver,
and by inverting the Bethe-Salpeter equation, we obtain $\Gamma^{imp}$.
For the the impurity model, the two-particle susceptibility is
formally obtained by differentiating the partition function 
\begin{equation}
\chi_{pp';rr'}^{imp}=\frac{1}{Z}\frac{\partial^2 Z}{\partial \Delta_{rp} \partial \Delta_{p'r'}}
\end{equation}
and resulting terms are sampled by the quantum Monte Carlo solver.
Notice that the susceptibility depends on four times, which can
be translated into three frequencies. It turns out that this
two-particle susceptibility can be broken up into two terms, of which
both depend on two frequencies only, which greatly reduces the
complexity of the problem. We sample these terms directly in frequency
during the Monte Carlo run, just like the single-particle Green's
function, to avoid the Fourier transform of a multidimensional object.

Once the impurity two-particle susceptibility $\chi_{pp';rr'}^{imp}$
is available, we compute the impurity polarization (bubble)
$\chi_{pp';rr'}^{0,imp}$, and extract
the particle-hole irreducible vertex $\Gamma_{pp';rr'}^{imp}$ by the
Bethe-Salpeter equation:
\begin{equation}
\chi_{pp'\nu;rr'\nu'}^{imp}=\chi_{pp'\nu;rr'\nu'}^{0,imp}+\chi_{pp'\nu;p_1p_1'\nu_1}^{0,imp}\Gamma_{p_1p_1'\nu_1;r_1r_1'\nu_1'}^{imp}\chi_{r_1r_1'\nu_1';rr'\nu'}^{imp}
\end{equation}
as 
\begin{equation}
\Gamma_{pp'\nu;rr'\nu'}^{imp}=(\chi^{0,imp})^{-1}_{pp'\nu;rr'\nu'}-(\chi^{imp})^{-1}_{pp'\nu;rr'\nu'}
\end{equation}
where $p$, $p'$, $r$, $r'$ are combined index of spin and orbital, and
$\nu$, $\nu'$ are fermionic Matsubara frequencies. These equations
depend also on the bosonic (center of mass) frequency $\Omega$, which
can be treated as an external parameter.

With the knowledge of the particle-hole irreducible vertex, we can compute
the non-local two particle susceptibility by the Bethe-Salpeter equation
\begin{equation}
\chi_{pp'\nu;rr'\nu'}^{q}=\chi_{pp'\nu;rr'\nu'}^{0,q}+\chi_{pp'\nu;p_1p_1'\nu_1}^{0,q}\Gamma_{p_1p_1'\nu_1;r_1r_1'\nu_1'}^{loc}\chi_{r_1r_1'\nu_1';rr'\nu'}^{q}
\end{equation}
as
\begin{equation}
\chi_{pp'\nu;rr'\nu'}^{q}=((\chi^{0,q})^{-1}-\Gamma^{loc})^{-1}_{pp'\nu;rr'\nu'}
\end{equation}
where $\chi_{pp'\nu;rr'\nu'}^{0,q}=-T\delta(\nu-\nu')\sum_{k} G_{k,rp}(i\nu)G_{k-q,p'r'}(i\nu-i\Omega)$
is the nonlocal one-particle bubble.
This ladder sum in the particle-hole channel incorporates most
important non-local spin and orbital fluctuations.

\begin{figure}[bht]
\includegraphics[width=0.7\linewidth]{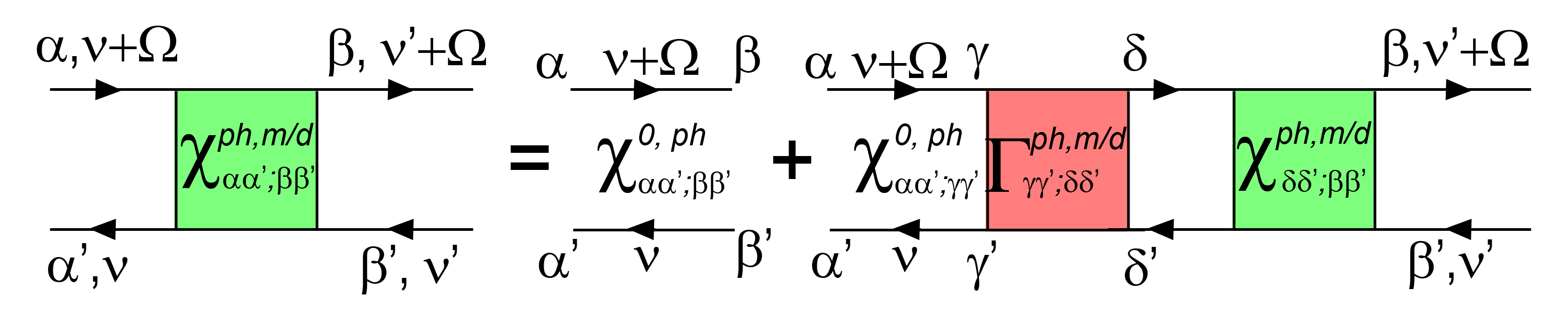}
\caption{ \textbf{Feynman diagram for computing the local and
    momentum-dependent two-particle susceptibility in the
    particle-hole spin (m) and charge (d) channels.}  Here the
  particle-hole irreducible vertex $\Gamma$ is assumed to be
  local. The momentum-dependent two-particle susceptibility is
  obtained by using the momentum-dependent one-particle bubble
  $\chi_q^0$.  The spin and charge susceptibility is computed by
  closing the corresponding two-particle susceptibility with the spin
  or charge bare vertex, and tracing over the internal indices, such as
  orbital ($\alpha$, $\alpha'$,$\beta$, $\beta'$) and frequency
  ($\nu$, $\nu'$).  }
\label{Chi_md}
\end{figure}

In the paramagnetic state (when the spin-orbit coupling is ignored),
there is a symmetry between the states with different spin, and the
equations can be block-diagonalized in spin. The following holds in
paramagnetic state $\chi_{\uparrow\uparrow;
  \uparrow\uparrow}=\chi_{\downarrow\downarrow;\downarrow\downarrow}$
and $\chi_{\uparrow\uparrow;
  \downarrow\downarrow}=\chi_{\downarrow\downarrow;\uparrow\uparrow}$
where the orbital index is omitted for simplicity. As a result, the
two-particle quantities can be expressed in terms of two independent
channels (no mixing between the two channels), the magnetic and the
charge channel as
\begin{equation}
\chi^{m}=\chi_{\uparrow\uparrow; \uparrow\uparrow}-\chi_{\uparrow\uparrow; \downarrow\downarrow}
\end{equation}
\begin{equation}
\Gamma^{m}=\Gamma_{\uparrow\uparrow; \uparrow\uparrow}-\Gamma_{\uparrow\uparrow; \downarrow\downarrow}
\end{equation}
\begin{equation}
\chi^{d}=\chi_{\uparrow\uparrow; \uparrow\uparrow}+\chi_{\uparrow\uparrow; \downarrow\downarrow}
\end{equation}
\begin{equation}
\Gamma^{d}=\Gamma_{\uparrow\uparrow; \uparrow\uparrow}+\Gamma_{\uparrow\uparrow; \downarrow\downarrow}
\end{equation}
Thus $\chi_{\uparrow\uparrow; \uparrow\uparrow}=(\chi^{d}+\chi^{m})/2$ and 
$\chi_{\uparrow\uparrow; \downarrow\downarrow}=(\chi^{d}-\chi^{m})/2$.
In addition, we have the following symmetry
\begin{equation}
<S^z(\tau)S^z(0)>=<S^+(\tau)S^-(0)>=<S^-(\tau)S^+(0)>
\end{equation} 
Accordingly we have
$\chi_{\downarrow\uparrow; \downarrow\uparrow}=\chi_{\uparrow\downarrow; \uparrow\downarrow}=\chi^{m}$.
Hence the spin index can be dropped in the paramagnetic state by using the above vertex in the magnetic 
and charge channel. 
As shown in Fig.\ref{Chi_md}, the two-particle vertex in the magnetic/charge (m/d) channel can be written as 
\begin{equation}
\chi_{\alpha \alpha'; \beta \beta'}^{m/d}(\nu, \nu')_{q, \Omega} =
((\chi^0)_{q,\Omega}^{-1}-\Gamma^{m/d})_{\alpha \alpha'; \beta \beta'}^{-1}(\nu, \nu')_{q,\Omega} 
\end{equation}
The spin or charge susceptibility $\chi(q, \omega)$ is obtained by
closing the corresponding nonlocal two-particle susceptibility
$\chi_{\alpha \alpha'; \beta \beta'}^{m/d}(\nu, \nu')_{q, \Omega}$
with the bare vertex $\mu$ and summation over the internal indices
\begin{equation}
\chi^{m/d}(q, \Omega)= 2 \sum_{\alpha, \alpha',\beta, \beta', \nu, \nu'} 
\mu_{\alpha \alpha'} \chi_{\alpha \alpha'; \beta \beta'}^{m/d}(\nu, \nu')_{q, \Omega} \mu_{\beta \beta'}
\end{equation}
Further computational details on magnetic susceptibility are available in Ref.~S\onlinecite{HPark-PRLS}. 

\subsection{Superconductivity}

\begin{figure}[bht]
\includegraphics[width=0.9\linewidth]{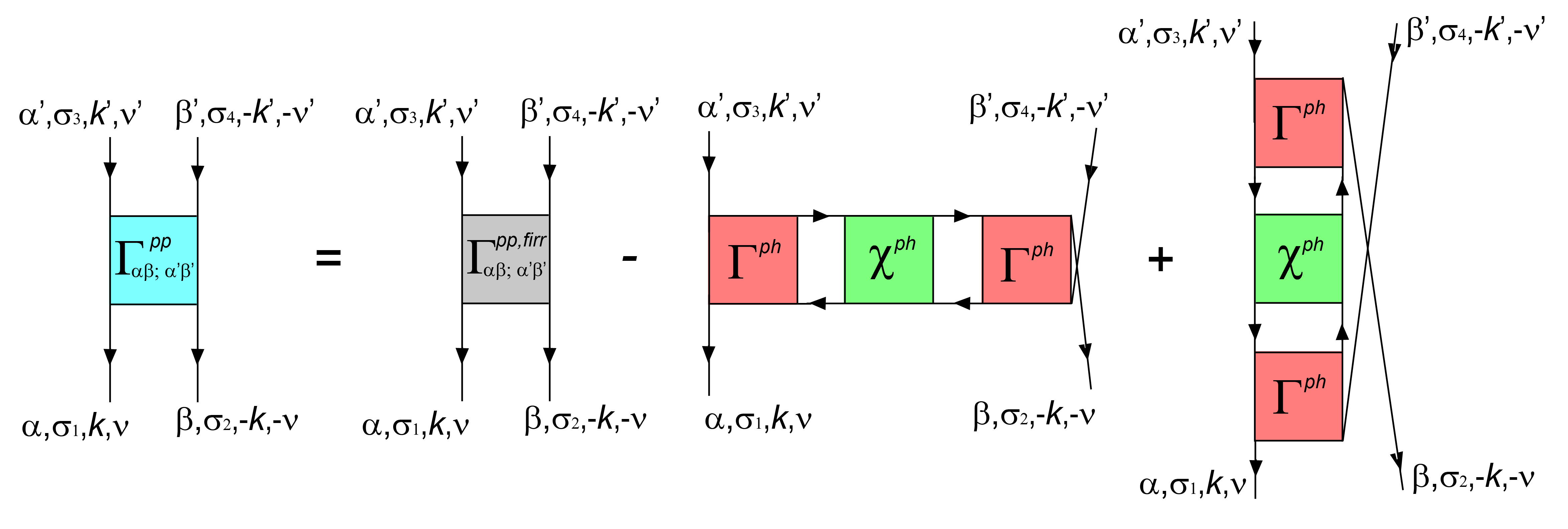}
\caption{
\textbf{Particle-particle irreducible vertex $\Gamma^{pp}_{\sigma_1\sigma_2;\sigma_3\sigma_4}$.}
It consists of fully-irreducible vertex $\Gamma^{pp,firr}$, and vertex which is reducible
in the particle-hole channel. There are two ways to arrange
particle-hole ladders, either horizontally ($\Gamma^{pp(1)}$) or
vertically ($\Gamma^{pp(2)}$), hence there
are two particle-hole contributions.
}
\label{Chi_pps1}
\end{figure}
A divergent susceptibility in the particle-particle channel signals
instability of the metallic state towards superconductivity.  To
obtain this susceptibility, we need to compute the particle-particle
irreducible vertex $\Gamma^{pp}$, depicted in Fig.\ref{Chi_pps1}.
It consists of the fully irreducible
vertex function $\Gamma^{firr}$ and the reducible vertex functions in
the particle-hole channels. There are two particle-hole channels,
because one can stack particle-hole ladders horizontally
(particle-hole channel 1) or vertically (particle-hole channel 2),
as shown in Fig.\ref{Chi_pps1}.
Notice that this equation contain all spin-fluctuation
diagrams~[S\onlinecite{Mazin-review}]. Indeed we recover the spin-fluctuation
theory if we replace $\Gamma^{ph}$ by constant number $U$, which is
treated as a phenomenological parameter in spin-fluctuation theory,
and propagators with free-electron Green's function. Since results are
very sensitive to the value and structure of this screened
interaction $\Gamma^{ph}$, it is important to determine it \textit{ab initio}.

\begin{figure}[bht]
\includegraphics[width=0.6\linewidth]{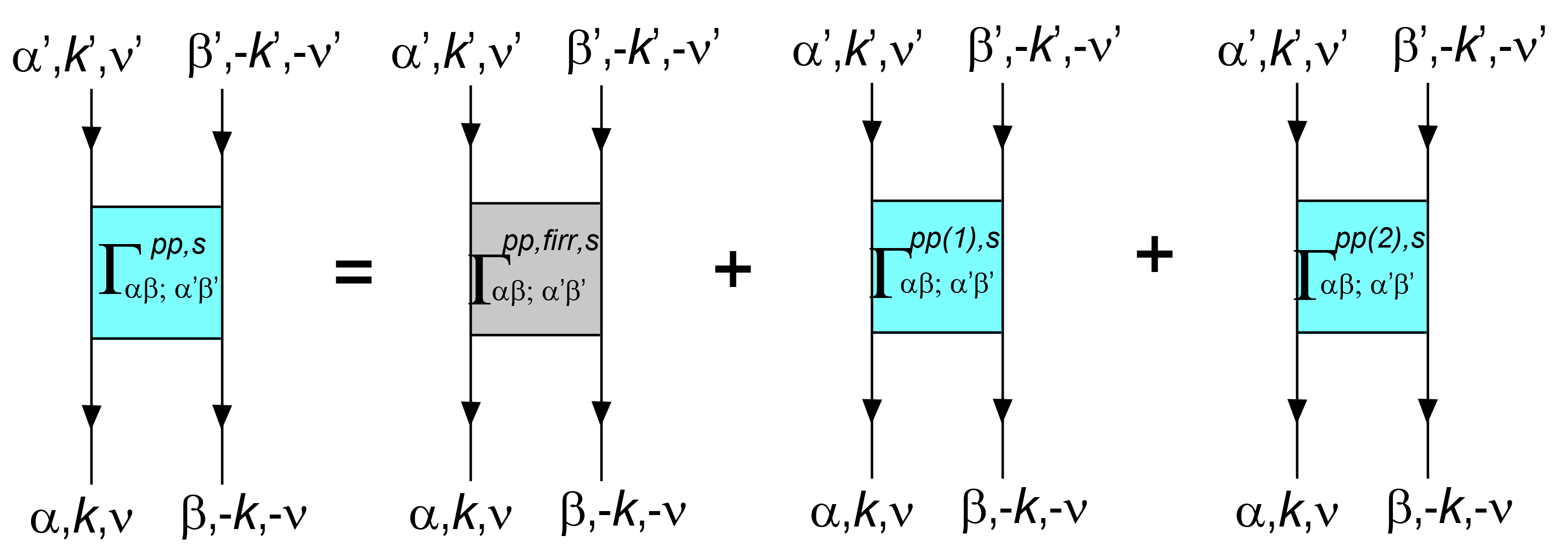}
\includegraphics[width=0.65\linewidth]{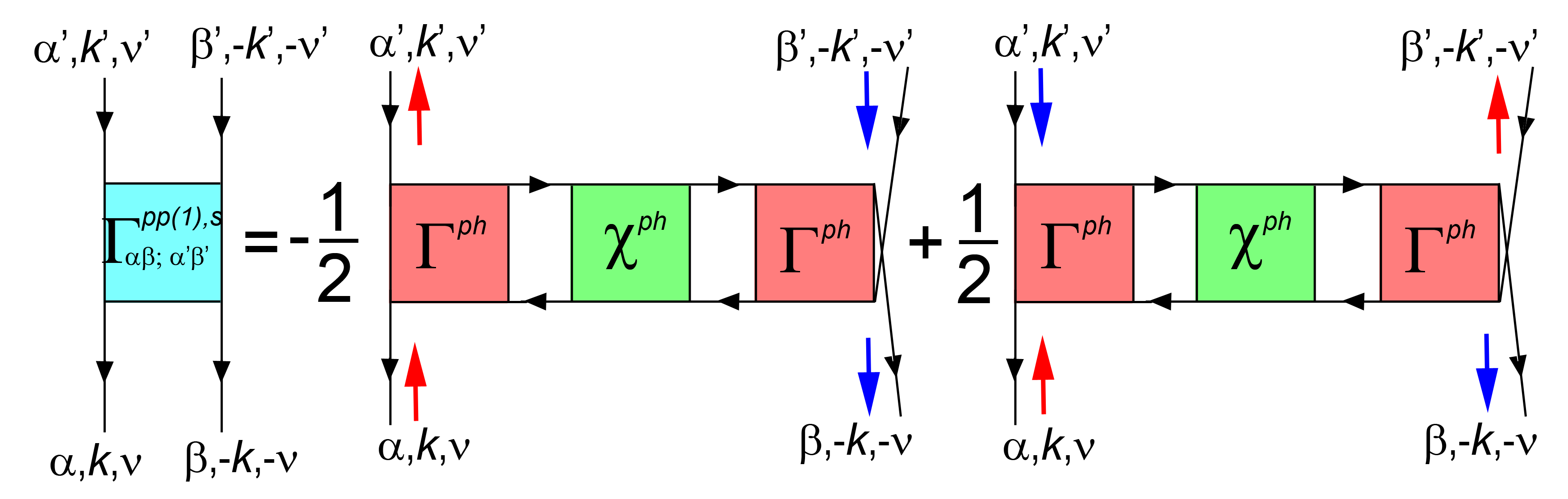}
\includegraphics[width=0.65\linewidth]{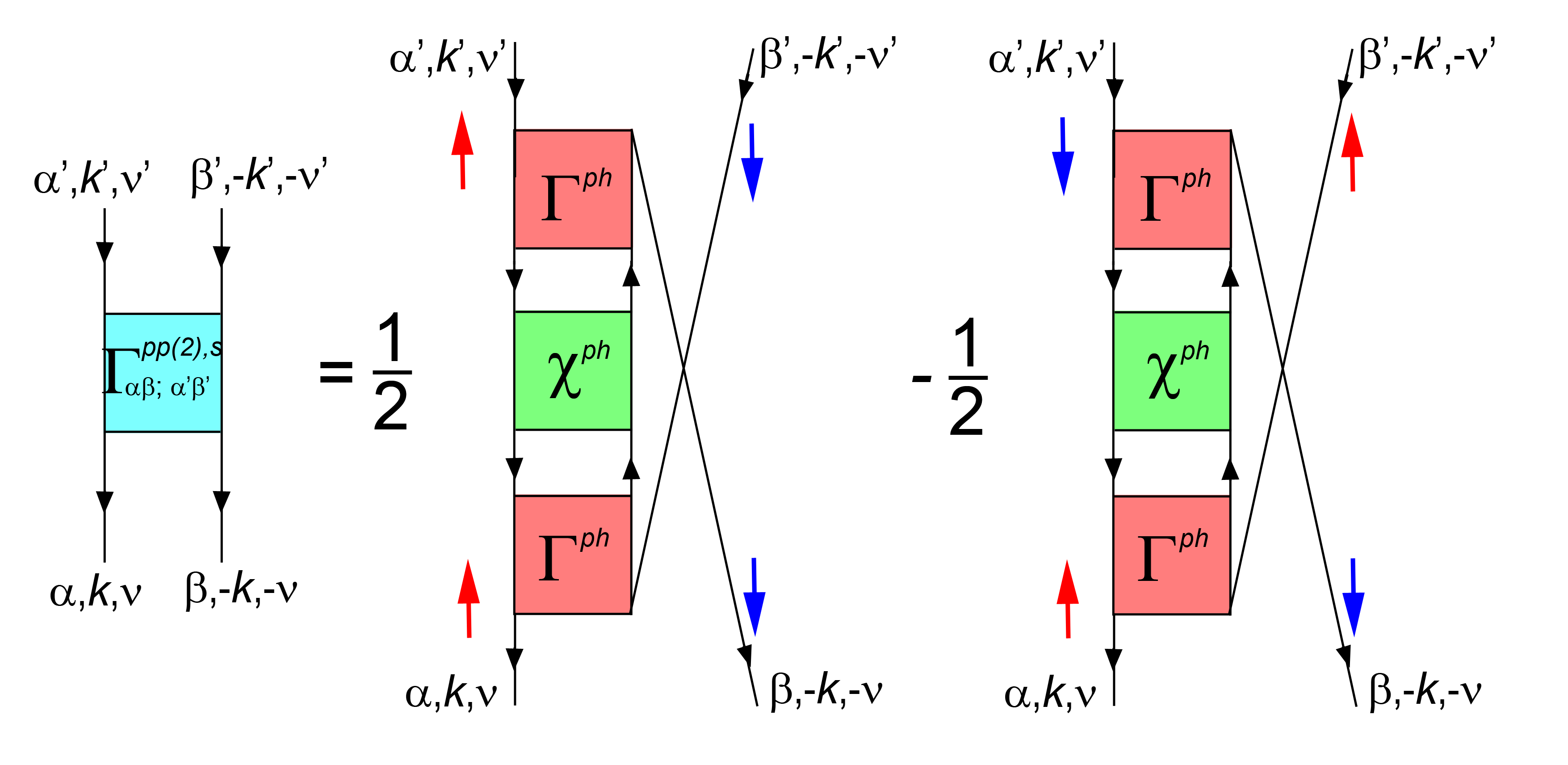}
\caption{
\textbf{Decomposition of particle-particle irreducible vertex for the spin singlet channel:}
We explicitly write the spin components, which contribute to the
vertex $\Gamma^{pp,s}$ in the singlet channel.
}
\label{Chi_pp1}
\end{figure}
In this report, we consider only the spin-singlet pairing and define
the singlet vertex $\Gamma^{pp,s}$ by
\begin{equation}
\Gamma^{pp,s}=\frac{1}{2}(\Gamma_{\uparrow\downarrow;\uparrow\downarrow}^{pp}-\Gamma_{\uparrow\downarrow;\downarrow\uparrow}^{pp})
\end{equation} 
For convenience, we rewrite $\Gamma^{pp,s}$ as the sum of the three
terms depicted in Fig.~\ref{Chi_pp1}:
\begin{equation}
\Gamma^{pp,s}=\Gamma^{pp,firr,s}+\Gamma^{pp(1),s}+\Gamma^{pp(2),s}.
\end{equation}
It then follows that the fully irreducible particle-particle vertex in
the spin singlet channel $\Gamma^{pp,firr,s}$ is
\begin{equation}
\Gamma^{pp,firr,s}=\frac{1}{2}(\Gamma^{pp,firr,s}_{\uparrow\downarrow;\uparrow\downarrow}-\Gamma^{pp,firr,s}_{\uparrow\downarrow;\downarrow\uparrow}),
\end{equation}

We can also express the rest of the objects in Fig.~\ref{Chi_pp1} in
terms of the above calculated particle-hole susceptibility $\chi^{ph}$ and
particle-hole irreducible vertex $\Gamma^{ph}$ by
\begin{equation}
\begin{aligned}
\Gamma^{pp(1),s} & =-\frac{1}{2}((\Gamma\chi\Gamma)_{\uparrow\uparrow;\downarrow\downarrow}^{ph}-(\Gamma\chi\Gamma)_{\downarrow\uparrow;\downarrow\uparrow}^{ph}) \\
                 & =\frac{3}{4}(\Gamma\chi\Gamma)^{ph, m}-\frac{1}{4}(\Gamma\chi\Gamma)^{ph, d}
\end{aligned}
\end{equation}
and
\begin{equation}
\begin{aligned}
\Gamma^{pp(2),s} & =\frac{1}{2}((\Gamma\chi\Gamma)_{\downarrow\uparrow;\downarrow\uparrow}^{ph}-(\Gamma\chi\Gamma)_{\uparrow\uparrow;\downarrow\downarrow}^{ph}) \\
                 & =\frac{3}{4}(\Gamma\chi\Gamma)^{ph, m}-\frac{1}{4}(\Gamma\chi\Gamma)^{ph, d}
\end{aligned}
\end{equation}
These diagrams are depicted in Fig.~\ref{Chi_pp1}. 

To simplify the notation, we then define
\begin{equation}
(\overline{\Gamma\chi\Gamma})^{ph} \equiv
  \frac{3}{4}(\Gamma\chi\Gamma)^{ph,m}-\frac{1}{4}(\Gamma\chi\Gamma)^{ph, d}.
\end{equation}
to connect the above computed magnetic and charge susceptibility/vertex with
the particle-particle irreducible vertex.
With this algebraic manipulation, we removed the need for the spin
index, however, we do have four orbital indices
($\alpha$,$\beta$,$\alpha'$,$\beta'$), three frequencies
(fermionic $\nu$,$\nu'$, and bosonic $\Omega$) and three momenta
($k$,$k'$,$q$) left. However, within DMFT the particle-hole
irreducible vertex $\Gamma^{ph}$ is local, therefore this quantity is
independent of momenta $k$ and $k'$ and can be written as
\begin{equation}
(\overline{\Gamma\chi\Gamma})^{ph}_{q,\Omega}(\alpha\beta,\nu;\alpha'\beta',\nu')
\end{equation}
Furthermore, the fully irreducible vertex within DMFT is local and hence
independent of momenta $\Gamma^{pp,firr,s}(\alpha \beta \nu;\alpha'\beta'\nu') $.

Finally, using these building blocks computed above, we explicitly write down the
irreducible particle-particle vertex
\begin{equation}
\begin{aligned}
& \Gamma^{pp,s}(\alpha \beta k\nu;\alpha'\beta'k'\nu')=
 \Gamma^{pp(1),s}(\alpha \beta k\nu;\alpha'\beta'k'\nu')  +\Gamma^{pp(2),s}(\alpha \beta k\nu;\alpha'\beta'k'\nu')+\Gamma^{pp,firr,s}(\alpha \beta \nu;\alpha'\beta'\nu') 
  \\
& = (\overline{\Gamma\chi\Gamma})^{ph}_{k'-k, \nu'-\nu}(\alpha'\alpha,\nu';\beta \beta',-\nu) 
 +(\overline{\Gamma\chi\Gamma})^{ph}_{-k'-k,-\nu'-\nu}(\beta'\alpha,-\nu';\beta \alpha',-\nu)+\Gamma^{pp,firr,s}(\alpha \beta \nu;\alpha'\beta'\nu')
\end{aligned}
\end{equation}
When the particle-particle ladder sum $\chi^{pp}=
((\chi^{0,pp})^{-1}-\Gamma^{pp})^{-1}$ is diverging, the normal state
is unstable to superconductivity. The sufficient condition is that
the matrix of $\Gamma^{pp}\chi^{0,pp}$ has an eigenvalue equal to
unity. The eigenvector with the largest eigenvalue gives the symmetry
of the superconducting order parameter. Explicitly, we are solving
the eigenvalue problem of the following matrix
\begin{equation}
\begin{aligned}
& -k_BT \sum_{k'\nu'\alpha'\beta' \gamma \delta} \Gamma^{pp,s}(\alpha \beta k\nu;\alpha'\beta'k'\nu') \chi^{0,pp}_{\alpha'\beta'\gamma\delta}(k'\nu') 
 \Delta_{\gamma\delta}(k'\nu')
=\lambda \Delta_{\alpha\beta}(k\nu)
\end{aligned}
\end{equation}
where the eigenvalue $\lambda$ is the pairing strength and the eigenfunction $\Delta$ is the pairing amplitude. 

To solve this eigenvalue problem, we make an approximation consistent
with the Bardeen-Cooper-Schrieffer (BCS) theory: since pairing occurs
at very low energy scale, we take the particle-particle vertex at zero
frequency as a proxy for its steep rise with lowering temperature, i.e.,
$\Gamma^{pp,s}(\alpha\beta k,\nu;\alpha'\beta'k',\nu')\approx \Gamma^{pp,s}(\alpha\beta k, 0^+;\alpha'\beta'k',0^+)$, 
hence we have 
\begin{equation}
\begin{aligned}
& -\sum_{k'\alpha'\beta'\gamma\delta} \Gamma^{pp,s}(\alpha\beta
  k0^+;\alpha'\beta'k'0^+)
 \left(k_BT\sum_{i\nu'}\chi^{0,pp}_{\alpha'\beta'\gamma\delta}(k',\nu')\right) 
 \Delta_{\gamma\delta}(k'0^+) 
=\lambda \Delta_{\alpha\beta}(k0^+)
\end{aligned}
\end{equation} 
where $k_BT
\sum_{\nu'}\chi^{0,pp}_{\alpha'\beta'\gamma\delta}(k',\nu')) =
k_BT \sum_{\nu'}G_{\alpha'\gamma}(k',\nu')G_{\beta'\delta}(-k',-\nu')$ 
is the one-particle bubble in the particle-particle channel.

To facilitate this calculation, we first transform the single-particle
Green's function to one-iron atom Brillouin zone. We then compute
pairing strength in orbital basis $\Delta_{\alpha\beta}(k)$ with above
formula, and we finally transform it to quasiparticle band basis by
so-called embedding method.
For DMFT calculation, we construct projector to
iron-$3d$ states $U_k$ (for details see Ref.~S\onlinecite{Haule-DMFT}),
which embeds the self-energy to the Kohn-Sham basis
($U_k \Sigma U_k^\dagger$) or projects Green's function expressed in the Kohn-Sham
basis to local basis ($U_k^\dagger G U_k$).
We then compute the eigenvectors and eigenvalues of the DMFT Green's
function at the Fermi level, defined by 
\begin{eqnarray}
(\varepsilon_k-\mu+ U_k \Sigma(\omega=0)U_k^\dagger) \psi^R(k) = \epsilon_k \psi^R(k)\\
\psi^L(k) (\varepsilon_k-\mu+ U_k \Sigma(\omega=0)U_k^\dagger)  = \epsilon_k \psi^L(k)
\end{eqnarray}
where $\epsilon_k$ are the DMFT eigenvalues at the Fermi level, which
determine the Fermi surface, and $\psi(k)$ are the eigenvectors.
We can then express the pairing strength on the Fermi surface as
\begin{equation}
\Delta_k \equiv \psi^L_k U_k \Delta (k) U_k^\dagger \psi^R_k
\end{equation}
Notice that $\Delta_k$ contains both the diagonal and off-diagonal
components in band basis. The latter are smaller, hence we plot
diagonal components in the manuscript $\Delta_{k,ii}$.


\begin{thebibliography}{10}


\bibitem{Stewart}Stewart, G. R.
Superconductivity in iron compounds.
\textit{Rev. Mod. Phys.} {\bf 83}, 1589-1652 (2011).

\bibitem{Hirschfeld}Hirschfeld, P.J., Korshunov, M.M., $\&$ Mazin, I.I.
Gap symmetry and structure of Fe-based superconductors.
\textit{Rep. Prog. Phys.} {\bf 74}, 124508 (2011).


\bibitem{PDai}Dai, P., Hu, J., $\&$ Dagotto, E.
Magnetism and its microscopic origin in iron-based high-temperature superconductors.
\textit{Nature Physics} {\bf 8}, 709-718 (2012).



\bibitem{LiFeAs}Borisenko,S. V. \textit{et al.}
One-Sign Order Parameter in Iron Based Superconductor.
\textit{Symmetry} {\bf 4}, 251-264 (2012).

\bibitem{LiFeAs2}Umezawa, K. \textit{et al.}
Unconventional Anisotropic s-Wave Superconducting Gaps of the LiFeAs Iron-Pnictide Superconductor.
\textit{Phys. Rev. Lett.} {\bf 108}, 037002 (2012).


\bibitem{Mazin_PRL}Mazin, I.I., Singh, D.J., Johannes, M.D., $\&$ Du, M. H. 
Unconventional Superconductivity with a Sign Reversal in the Order Parameter of LaFeAsO$_{1-x}$F$_x$.
\textit{Phys. Rev. Lett.} {\bf 101}, 057003 (2008).


\bibitem{review}Kotliar, G. \textit{et al.}
Electronic structure  calculations with dynamical mean-field  theory.
\textit{Rev. Mod. Phys.} \textbf{78}, 865-951 (2006).

\bibitem{Yin-NM}Yin, Z.P., Haule, K., $\&$ Kotliar, G.
Kinetic frustration and the nature of the magnetic and paramagnetic states in iron pnictides and iron chalcogenides.
\textit{Nature Materials} {\bf 10}, 932-935 (2011).

\bibitem{Yin-PRB}Yin, Z.P., Haule, K., $\&$ Kotliar, G.
Fractional power-law behavior and its origin in iron-chalcogenide and ruthenate superconductors: Insights from first-principles calculations.
\textit{Phys. Rev. B} {\bf 86}, 195141 (2012).

\bibitem{Yin-NP}Yin, Z.P., Haule, K., $\&$ Kotliar, G.
Magnetism and Charge Dynamics in Iron Pnictides.
\textit{Nature Physics} {\bf 7}, 294-297 (2011).

\bibitem{HPark-PRL}Park, H., Haule, K., $\&$ Kotliar, G.
Magnetic Excitation Spectra in BaFe$_2$As$_2$: A Two-Particle Approach within a Combination of the Density Functional Theory and the Dynamical Mean-Field Theory Method.
\textit{Phys. Rev. Lett.} {\bf 107}, 137007 (2011).

\bibitem{Kuroki}Kuroki, K. \textit{et al.}
Unconventional Pairing Originating from the Disconnected Fermi Surfaces of Superconducting LaFeAsO$_{1-x}$F$_x$.
\textit{Phys. Rev. Lett.} {\bf 101}, 087004 (2008).

\bibitem{Kreisel}Kreisel, A. \textit{et al.}
Spin fluctuations and superconductivity in K$_x$Fe$_{2-y}$Se$_2$.
\textit{Phys. Rev. B} {\bf 88}, 094522 (2013).

\bibitem{Onari}Onari, S. \textit{et al.}
$S$-wave superconductivity due to orbital and spin fluctuations in Fe-pnictides: self-consistent vertex correction with self-energy (SC-VC$\Sigma$) analysis.
arXiv:1307.6119.

\bibitem{FWang}Wang, F., Zhai, H., Ran, Y., Vishwanath, A., $\&$ Lee, D.-H.
Functional Renormalization-Group Study of the Pairing Symmetry and Pairing Mechanism of the FeAs-Based High-Temperature Superconductor.
\textit{Phys. Rev. Lett.} {\bf 102}, 047005 (2009).

\bibitem{Qimiao}Yu, R., Zhu, J.-X., $\&$ Si, Q.
Orbital-selective superconductivity, gap anisotropy and spin resonance excitations in a multiorbital t-J1-J2 model for iron pnictides.
arXiv:1306.4184.

\bibitem{chi_BFA}Harriger, L.W. \textit{et al.}
Nematic spin fluid in the tetragonal phase of BaFe$_2$As$_2$.
\textit{Phys. Rev. B} {\bf 84}, 054544 (2011).

\bibitem{chi_LiFeAs}Wang, M. \textit{et al.}
Antiferromagnetic spin excitations in single crystals of nonsuperconducting Li$_{1-x}$FeAs.
\textit{Phys. Rev. B} {\bf 83}, 220515(R) (2011).

\bibitem{chi_BFNA}Liu, M.S. \textit{et al.}
Nature of magnetic excitations in superconducting BaFe$_{1.9}$Ni$_{0.1}$As$_{2}$.
\textit{Nature Physics} {\bf 8}, 376-381 (2012).

\bibitem{chi_BKFA}Wang, M. \textit{et al.}
A magnetic origin for high temperature superconductivity in iron pnictides.
arXiv:1303.7339.

\bibitem{HBRhee}Rhee, H.B. $\&$ Pickett, W.E.
Contrast of LiFeAs with isostructural, isoelectronic, and non-superconducting MgFeGe.
\textit{J. Phys. Soc. Jpn.} {\bf 82}, 034714 (2013).


\bibitem{MgFeGe-Mazin}Jeschke, H.O., Mazin, I.I., $\&$ Valenti, R. 
Why MgFeGe is not a superconductor.
arXiv:1305.7368.

\bibitem{moment-FeTe}Bao, W. {\it et al.} 
Incommensurate magnetic order in the $\alpha$-Fe(Te,Se) superconductor systems.
\textit{Phys. Rev. Lett.} {\bf 102}, 247001 (2009).


\bibitem{TMoriya}Moriya, T., Takahashi, Y., $\&$ Ueda, K.
Antiferromagnetic spin fluctuations and superconductivity in two-dimensional metals-a possible model for high Tc oxides.
\textit{J. Phys. Soc. Jpn.} {\bf 59}, 2905-2015 (1990).

\bibitem{DPines}Monthoux, P., Balatsky, A. V., $\&$ Pines, D
Toward a theory of high temperature superconductivity in the antiferromagnetically correlated cuprate oxide.
\textit{Phys. Rev. Lett.} {\bf 67}, 3448-3452 (1991).

\bibitem{Note-KFA}In Ref.\onlinecite{Yin-NM}, the nominal valence of Fe, as
  needed for double counting correction, was fixed to 3$d^6$ across
  all Fe compounds. To improve agreement with experiment, it is better
  to use the nominal valence of each compound, as has been done in the
  current paper, for example 3$d^{5.5}$ for Fe in KFe$_2$As$_2$.  This
  results in a substantially larger mass enhancement (a factor of two
  for the $xy$ orbital) in KFe$_2$As$_2$ compared to the previous
  study in Ref.\onlinecite{Yin-NM}.


\bibitem{KFA-SQW}Lee, C.H. \textit{et al.}
Incommensurate spin fluctuations in hole-overdoped superconductor KFe$_2$As$_2$.
\textit{Phys. Rev. Lett.} 106, 067003 (2011).


\bibitem{KFS-moment}Bao, W. {\it et al.}
A Novel Large Moment Antiferromagnetic Order in K$_{0.8}$Fe$_{1.6}$Se$_2$ Superconductor.
\textit{Chinese Phys. Lett.} {\bf 28}, 086104 (2011).


\bibitem{Hong-Ding}Ding, H. {\it et al.}, private communication.

\end{thebibliography}

\begin{thebibliography}{10}
  
\bibitem{DMFT-RMP2006}Kotliar, G.,  Savrasov, S. Y., Haule, K., Oudovenko, V. S., Parcollet, O. $\&$ Marianetti, C. A.
Electronic structure  calculations with dynamical mean-field  theory.
\textit{Rev. Mod. Phys.} \textbf{78}, 865-951 (2006).

\bibitem{Haule-DMFT}Haule, K., Yee, C.-H. $\&$ Kim, K.
Dynamical mean-field theory within the full-potential methods: electronic structure of CeIrIn$_5$, CeCoIn$_5$, and CeRhIn$_5$.
\textit{Phys. Rev. B} {\bf 81}, 195107 (2010).

\bibitem{wien2k}Blaha, P.,
Schwarz, K., Madsen, G. K. H., Kvasnicka, D. $\&$ Luitz, J.
\textsf{WIEN2K}
An augmented plane wave $+$ local orbitals program for calculating crystal properties,
(K. Schwarz, Techn. Univ. Wien, Austria, 2001).


\bibitem{HPark-PRLS}Park, H., Haule, K., $\&$ Kotliar, G.
Magnetic Excitation Spectra in BaFe$_2$As$_2$: A Two-Particle Approach within a Combination of the Density Functional Theory and the Dynamical Mean-Field Theory Method.
\textit{Phys. Rev. Lett.} {\bf 107}, 137007 (2011).

\bibitem{Haule-QMC}Haule, K.
Quantum Monte Carlo impurity solver for cluster dynamical mean-field theory and electronic structure calculations with adjustable cluster base.  
\textit{Phys. Rev. B} \textbf{75}, 155113 (2007).

\bibitem{Werner}Werner, P.,  Comanac, A.,  de Medici, L.,  Troyer, M. $\&$ Millis, A. J. 
Continuous-time solver for quantum impurity models.
\textit{Phys. Rev. Lett.} \textbf{97}, 076405 (2006).


\bibitem{Yin-NMS}Yin, Z.P., Haule, K., $\&$ Kotliar, G.
Kinetic frustration and the nature of the magnetic and paramagnetic states in iron pnictides and iron chalcogenides.
\textit{Nature Materials} {\bf 10}, 932-935 (2011).

  
\bibitem{a-FeTe}Tropeano, M., Pallecchi, I., Cimberle, M. R., Ferdeghini, C., Lamura, G., Vignolo, A., Martinelli, A., Palenzona, A. $\&$ Putti, M. 
Transport and superconducting properties of Fe-based superconductors: SmFeAs(O$_{1-x}$F$_x$) versus Fe$_{1+y}$(Te$_{1-x}$Se$_x$).
\textit{Supercond. Sci. Techolol.} {\bf 23}, 054001 (2010). 

\bibitem{a-FeTe-2}Martinelli, A., Palenzona, A., Tropeano, M., Ferdeghini, C., Putti, M., Cimberle, M. R., Nguyen, T. D., Affronte, A. $\&$ Ritter, C.
From antiferromagnetism to superconductivity in Fe$_{1+y}$(Te$_{1-x}$,Se$_x$) (0$<$$x$$<$0.20): a neutron powder diffraction analysis. 
\textit{Phys. Rev. B} {\bf 81}, 094115 (2010).

\bibitem{Guo}Guo, J. G., Jin, S. F., Wang, G., Wang, S. C., Zhu, K. X., Zhou, T. T., He, M. $\&$ Chen, X. L.  
Superconductivity in the iron selenide K$_x$Fe$_2$Se$_2$ (0$\leq x \leq$1.0).
\textit{Phys. Rev. B} {\bf 82}, 180520(R) (2010).


\bibitem{a-FeSe}Phelan, D., Millican, J. N., Thomas, E. L., Leao, J. B., Qiu, Y. $\&$ Paul, P.
Neutron scattering Measurements of the phonon density of states of FeSe$_{1-x}$ superconductors.
\textit{Phys. Rev. B} {\bf 79}, 014519 (2009).

\bibitem{a-LiFeAs}Tapp, J. H., Tang, Z. J., Lv, B., Sasmal, K., Lorenz, B., Chu, P. C. W. $\&$ M. Guloy, A. M. 
LiFeAs: an intrinsic FeAs-based superconductor with Tc = 18K.
\textit{Phys. Rev. B} {\bf 78}, 060505(R) (2008).


\bibitem{a-K122}Johrendt, D. $\&$ Poettgen, R.
Superconductivity, magnetism and crystal chemistry of Ba$_{1-x}$K$_x$Fe$_2$As$_2$.
\textit{Physica C} {\bf 469}, 332-339 (2009).

\bibitem{a-Ba122}Rotter, M., Tegel, M., Johrendt, D., Schellenberg, I., Hermes, W.,  $\&$ P\"{o}ttgen, R.
Spin-density-wave anomaly at 140 K in the ternary iron arsenide BaFe$_2$As$_2$.
\textit{Phys. Rev. B} {\bf 78}, 020503(R) (2008).


\bibitem{a-LaFeAsO}Kamihara, Y., Watanabe, T., Hirano, M. $\&$ Hosono, H.
Iron-based layered superconductor La[O$_{1-x}$F$_x$]FeAs ($x$ = 0.05-0.12) with Tc = 26 K.
\textit{J. Amer. Chem. Soc.} {\bf 130}, 3296-3297 (2008).


\bibitem{a-Sr122}Tegel, M.,  Rotter, M.,  Weiss, V., Schappacher, F. M., Poettgen, R. $\&$ Johrendt, D.
Structural and magnetic phase transitions in the ternary iron arsenides SrFe$_2$As$_2$ and EuFe$_2$As$_2$.
\textit{J. Phys.: Condens. Matter} {\bf 20}, 452201 (2008).

\bibitem{a-LaFePO}Kamihara, Y., Hiramatsu, H., Hirano, M., Kawamura, R., Yanagi, H., Kamiya, T. $\&$ Hosono, H.
Iron-based layered superconductor: LaOFeP.
\textit{J. Am. Chem. Soc.} {\bf 128}, 10012-10013 (2006).

\bibitem{a-LiFeP}Deng, Z., Wang, X.C., Liu, Q.Q., Zhang, S.J., Lv, Y.X., Zhu, J.L., Yu, R.C., $\&$ Jin, C.Q. 
A new 111 type iron pnictide superconductor LiFeP.
\textit{Europhys. Lett.} {\bf 87}, 3704 (2009).

\bibitem{a-BaFe2P2}Analytis, J.G., Chi, J.-H., McDonald, R.D., Riggs, S.C., $\&$ Fisher, I.R.
Enhanced Fermi surface nesting in superconducting BaFe$_2$(As$_{1-x}$P$_x$)$_2$ revealed by de Haas-van Alphen effect.
\textit{Phys. Rev. Lett.} {\bf 105}, 207004 (2010).

\bibitem{Kutepov}Kutepov, A., Haule, K., Savrasov, S. Y. $\&$ Kotliar, G.
Self consistent GW determination of the interaction strength: application to the iron arsenide superconductors.
\textit{Phys. Rev. B} {\bf 82}, 045105 (2010).




\bibitem{Yin-np}Yin, Z. P., Haule, K. $\&$ Kotliar, G.
Magnetism and charge dynamics in iron pnictides.
\textit{Nat. Phys.} {\bf 7}, 294-297 (2011).

\bibitem{review1996}Georges, A., Kotliar, G., Krauth, W., $\&$ Rozenberg, M.J.
Dynamical mean-field theory of strongly correlated fermion systems and the limit of infinite dimensions.
\textit{Rev. Mod. Phys.} {\bf 68}, 13-125 (1996).


\bibitem{Mazin-review}Hirschfeld, P.J., Korshunov, M.M., $\&$ Mazin, I.I.
Gap symmetry and structure of Fe-based superconductors.
\textit{Rep. Prog. Phys.} {\bf 74}, 124508 (2011).


\end{thebibliography}
\end{document}